# OS4M: Achieving Global Load Balance of MapReduce Workload by Scheduling at the Operation Level

Liya Fan[1], Bo Gao[1], Fa Zhang[2], Zhiyong Liu[2]
1 IBM China Research Laboratory
2 Institute of Computing Technology, Chinese Academy of Sciences
Beijing China
fanliya@cn.ibm.com, tju_gb@163.com, fazhang@gmail.com, zyliu@ict.ac.cn

*Abstract*—The efficiency of MapReduce is closely related to its load balance. Existing works on MapReduce load balance focus on coarse-grained scheduling. This study concerns fine-grained scheduling on MapReduce operations, with each operation representing one invocation of the Map or Reduce function. By default, MapReduce adopts the hash-based method to schedule Reduce operations, which often leads to poor load balance. In addition, the copy phase of Reduce tasks overlaps with Map tasks, which significantly hinders the progress of Map tasks due to I/O contention. Moreover, the three phases of Reduce tasks run in sequence, while consuming different resources, thereby under-utilizing resources. To overcome these problems, we introduce a set of mechanisms named OS4M (Operation Scheduling for MapReduce) to improve MapReduce's performance. OS4M achieves load balance by collecting statistics of all Map operations, and calculates a globally optimal schedule to distribute Reduce operations. With OS4M, the copy phase of Reduce tasks no longer overlaps with Map tasks, and the three phases of Reduce tasks are pipelined based on their operation loads. OS4M has been transparently incorporated into MapReduce. Evaluations on standard benchmarks show that OS4M's job duration can be shortened by up to 42%, compared with a baseline of Hadoop.

## 1 Introduction

MapReduce [DG04] has emerged as a powerful computing framework for processing big data in Cloud and distributed computing. It processes data with two basic functions: Map and Reduce. MapReduce works by partitioning the workload into a set of Map/Reduce tasks. Each Map/Reduce task consists of one or more Map/Reduce operations. *In particular, each invocation of the Map/Reduce function is named a Map/Reduce operation.* These operations are distributed across available Map/Reduce task slots and executed in parallel. Therefore, achieving load balance for MapReduce is critical for high parallel efficiency, high resource utilization, small job duration, etc. Previous works focus on coarse-grained scheduling of MapReduce workload. For example, [K+11] [SL10] [B+05] and [BD11] address the load balance problem at the job or task level. In this study, we consider the problem at a fine-grained level: the operation level.

*For simplicity, we define the load of a task or operation as the number of key-value pairs to be processed by it.* Please note that the effectiveness of our approach does not depend on this definition. We will show that achieving load balance for MapReduce operations is difficult, especially for Reduce operations. By default, current MapReduce adopts the hash based method to schedule Reduce operations. In particular, input to the hash function is the key related to the operation, and the output is the task ID. This method achieves load balance only when each operation has roughly equal load, and output of the hash function is evenly distributed in its range. However, these assumptions are often far from reality [K+11b]. For example, Fig. 1 shows the results obtained by default MapReduce on a PUMA benchmark (Purdue MapReduce benchmarks suite) [T+12] with 10GB input data. Fig 1(a) gives the cumulative distribution function of Reduce operation loads. The smallest operation has only one pair, while the largest has $1.97 \times 10^6$. From Fig. 1(b), it can be seen that task loads produced by such skewed operations are also skewed. The largest task load is 2.82 times larger than the smallest.

The second problem for current MapReduce is that, Reduce tasks starts to copy Map outputs immediate after the first Map task is finished. This will lead to I/O contention between Reduce tasks and subsequent Map tasks. Fig. 2 illustrates this by a job running a PUMA [T+12] benchmark. It can be observed that the first wave of Map tasks (the first 40%) took 45 seconds, while the second wave (from 40% to 80%) took 86 seconds. After that, as more Map outputs are generated, the contention between Map and Reduce tasks becomes more intensive, making the last wave of Map tasks (from 80% to 100%) extremely slow. This problem significantly hinders the overall job progress, since Reduce functions cannot run until the last Map task is finished [W+11].

The last problem is related to the three phases of Reduce tasks: copy, sort, and run. The three phases mainly consume different resources. In general, the copy phase is network I/O intensive, the sort phase is disk I/O intensive, and the run phase is CPU intensive. Current MapReduce processes the three phases in sequence, leading to low resource utilization. In addition, this design also enlarges the barrier between Map and Reduce phases [L+11], because the Reduce function cannot be invoked until all of the tasks' operations have been copied and sorted.

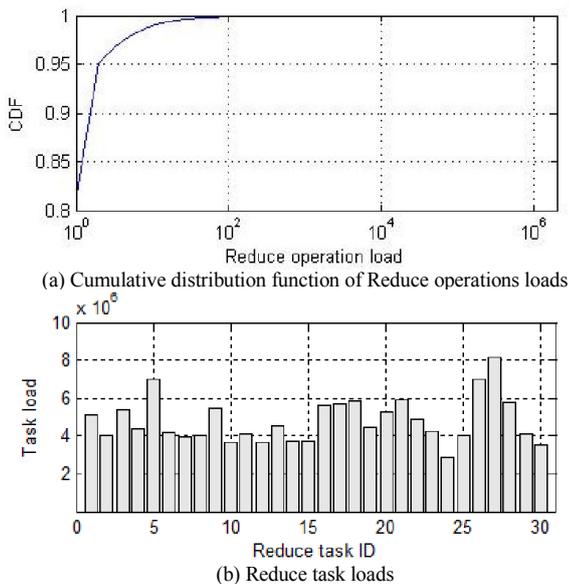

(a) Cumulative distribution function of Reduce operations loads

(b) Reduce task loads

Figure 1. Reduce Operation/Task loads produced by default MapReduce. The benchmark is Ranked Inverted Index (RII) from PUMA, with 10 GB input data. The MapReduce implementation is Hadoop 1.0.4.

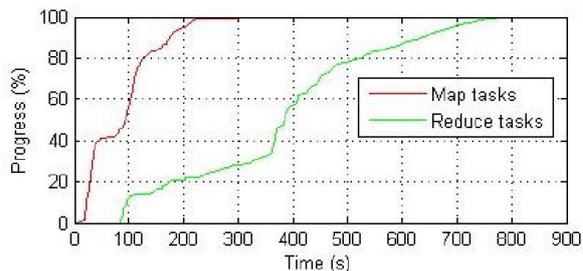

Figure 2. Task progress of a job running Inverted Index (II), with 5GB input data. The MapReduce implementation is Hadoop 1.0.4.

To solve these problems, we design and implement a set of mechanisms named OS4M (Operation Scheduling for MapReduce). OS4M achieves load balance by scheduling at the level of operations. To obtain a globally optimal schedule, the load of each operation should be known [G+12]. This is done by extending MapReduce's communication mechanism to collect operation statistics. With OS4M, the copy phase of Reduce tasks starts after all Map tasks are finished. Although this may cause some delay for Reduce tasks, Map tasks progress much faster, because the I/O contention between Map and Reduce tasks is avoided. Therefore, the overall progress is also faster. For OS4M, the three phases of Reduce tasks are pipelined to increase resource utilization. The order of operations on the pipeline is determined by their loads in order to minimize the barrier between Map and Reduce phases. In addition, the users can customize the granularity of the pipeline.

We have transparently incorporated OS4M into Hadoop [Wh12], the most widely-used implementation of MapReduce. Evaluations on PUMA benchmarks [T+12] show that OS4M achieves better load balance and shorter task durations. The performance gain for the whole job can be as high as 42%.

Contributions of this paper include: 1) We introduce a framework for fine-grained scheduling of MapReduce workload. To the best of our knowledge, this is the first attempt to improve the scheduling of MapReduce workload at the operation level. 2) We design a communication mechanism to collect operation statistics. By summarizing operation statistics we obtain the key distribution of intermediate pairs, which is the input of our operation scheduling algorithm. This mechanism can also be used to collect other statistics for conducting other enhancements to MapReduce. 3) We design a pipeline for Reduce workloads. The pipeline greatly increases resource utilization, as well as decreases task duration. 4) We formulate the load balance problem of Reduce operations as a scheduling problem denoted as $P\|C_{max}$ [Ho98]. We design an algorithm to solve it based on the key distribution of intermediate pairs. The algorithm is fast and produces near-optimal schedules for general MapReduce jobs.

The rest of this paper is organized as follows: Section 2 gives the background knowledge. Section 3 analyzes the load balance problem and formalizes it to $P\|C_{max}$. Section 4 introduces OS4M's enhancements to MapReduce. Section 5 gives evaluation results. Section 6 discuses problems related to the implementation and evaluation of OS4M. We discuss related work in Section 7 and conclude in Section 8.

## 2 BACKGROUND

An important notation for MapReduce is *job*. It represents all the work that is done after one submits his computation request and before he gets his result (see Fig. 3). A job contains a number of *tasks*, and a task contains one or more *operations*. Hardware resources for MapReduce workload are abstracted as a pool of task *slots*. A task slot can be either a Map task slot or a Reduce task slot. At any time, each task slot can process at most one task.

A general MapReduce job contains two types of tasks: Map tasks and Reduce tasks. The inputs of Map tasks are a set of input key-value pairs. These input pairs are split into a number of subsets, and each subset is processed by a Map task and a Map operation[1]. These tasks are distributed across all Map task slots and conducted in parallel. Each Map operation processes its input key-value pairs, and generates some intermediate key-value pairs.

The inputs of Reduce tasks are the intermediate key-value pairs produced by Map tasks. Similarly, the inputs are distributed to Reduce tasks, and processed in Reduce task slots in parallel. For Reduce tasks, however, there is a problem of deciding which intermediate pairs are processed by which Reduce tasks. This decision is called *partitioning*. Partitioning must comply with the following constraint: all intermediate pairs produced by all Map tasks with the same key are processed by the same Reduce operation in the same Reduce task. This constraint is the key difference between Reduce and Map operations. We call it *Reduce Input Constraint*. Another difference from Map tasks is that a Reduce task usually contains more than one Reduce

---

[1] Since each Map task contains exactly one Map operation, the terms Map task and Map operation can be used interchangeably.

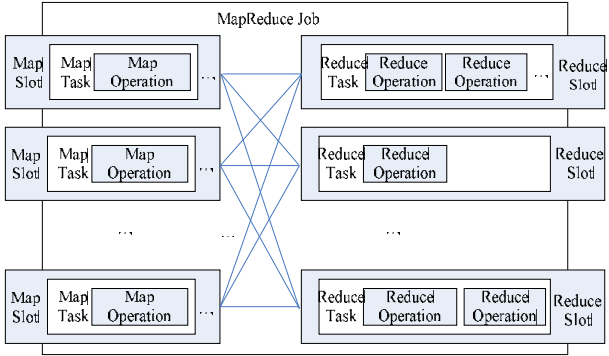

Figure 3. The Basic Structure of a MapReduce Job

operation (see Fig. 3), with each operation processing intermediate pairs with a particular key.

Because of the Reduce Input Constraint, intermediate pairs cannot be distributed to Reduce operations at will. This is the fundamental reason for the load unbalance of Reduce operations and Reduce tasks.

## 3 THE LOAD BALANCE PROBLEM OF MAPREDUCE OPERATIONS

This section discusses the load balance problems for Map and Reduce operations separately. Our objective is to balance the loads of different Map/Reduce task slots. Since the most widely used implementation of MapReduce, Hadoop, is primarily designed and optimized for homogeneous clusters [Z+08] [WK11] [A+12], we assume homogeneous nodes and homogeneous task slots. Moving OS4M to the heterogeneous environment is left as our future work.

For simplicity, we assume the numbers of Map and Reduce tasks slots are both equal to $m$. This is consistent with the default settings of Hadoop [Wh12]. However, the effectiveness of OS4M does not rely on this assumption.

### 3.1 Load Balance for Map Operations

Since the Map operation does not have any constraints like the Reduce Input Constraint, we can split the set of input key-value pairs at will. Therefore load balance for Map operations is straightforward. Basically, the input key-value pairs can be split into $m$ subsets of equal size, corresponding to $m$ Map operations. Alternatively, the input pairs can be split into $wm$ equal-sized subsets ($w$ = 1, 2, …), corresponding to $wm$ Map operations, so each Map task slot processes $w$ Map operations. Anyway, the total number of input key-value pairs to process is the same for all Map task slots.

$w$ is called the number of *waves* for Map tasks. A larger value of $w$ results in smaller operations. This generally leads to better load balance, but more initialization and cleaning cost. On the other hand, a smaller value of $w$ leads to worse load balance, but less initialization and cleaning cost. Therefore, the key of scheduling Map operations is not the scheduling algorithm, but selecting a proper value of $w$ to achieve the optimal trade-off. In this paper, we do not focus on the scheduling of Map operations.

### 3.2 Load Balance for Reduce Operations

As discussed earlier, a Reduce task may contain more than one Reduce operations. In other words, the Reduce operations on a Reduce task slot can be organized into a number of Reduce tasks. In order to minimize the initialization and cleaning up costs, OS4M organizes all Reduce operations on a Reduce task slot into a single Reduce task. This is also typical for default MapReduce [A+12]. Therefore, in the following discussion, the terms Reduce task and Reduce task slot can be used interchangeably.

Scheduling Reduce operations is difficult, due to the Reduce Input Constraint. This constraint forbids splitting the set of intermediate key-value pairs at will. To achieve load balance, the split must depend on the number of intermediate pairs with each intermediate key [G+12], and we call it the *key distribution of intermediate pairs*. However, this distribution relies on the specific input, and cannot be obtained before all Map operations are finished.

*For brevity, for the following discussion, we use pair to refer to intermediate pair, and key to refer to intermediate key.* According to Section 2, all pairs with a certain key are processed by a single Reduce operation, so the number of these pairs decides the load of the Reduce operation. In most cases, the numbers of pairs with different keys are quite different. Therefore the loads of Reduce operations differ greatly. Fig. 1(a) gives an example from our experiments. In the job, the largest operation has nearly two million pairs, while the smallest has only one. It is a challenging problem to achieve load balance in the presence of operations with highly varying loads. As we will see, the problem is strongly NP-hard.

Because of this difficulty, current MapReduce adopts simple scheduling strategies. Specifically, a pair $<k, v>$ is processed by the $i$th Reduce task [DG04], where

$$i = (|\text{Hash}(k)| \mod m) + 1 \quad (3\text{-}1)$$

*Hash* is a hash function for the key $k$. That means, the Reduce operation corresponding to key $k$ is processed by the $i$th Reduce task. It can be verified that this method complies with the Reduce Input Constraint, but the consequent load balance is poor, because the essence of such scheduling strategy is randomly selecting a task for each operation. Fig. 1(b) gives an example from our experiment. It can be seen that the load balance produced by default MapReduce is poor.

The failure of MapReduce's default scheduling strategy lies in its following implicit assumptions:
1) The key distribution of intermediate pairs is uniform.
2) The outputs of the hash function are uniformly distributed in its range.

It can be seen that when both assumptions hold, MapReduce's default scheduling strategy achieves load balance. However, these assumptions are often far from reality (For example, Fig 1(a) gives a counter-example of assumption 1).

To formulate the problem for scheduling Reduce operations, we suppose there are totally $n$ different keys, and the number of pairs with the $j$th key is $k_j$ ($j$ = 1, 2, …, $n$) (See

TABLE 1. MAIN SYMBOLS USED IN ANALYSIS

| Symbol | Description |
| --- | --- |
| $n$ | The number of operations (clusters) |
| $m$ | The number of Map/Reduce task slots |
| $M$ | The number of Map tasks (operations) |
| $r$ | The number of Reduce tasks |
| $w$ | The number of waves for Map operations. |
| $t$ | The number of TaskTrackers |
| $k_j$ | The number of pairs in the $j$th Reduce operation (cluster) |
| $p_i$ | The number of pairs processed by the $i$th Reduce task. |

Table 1 for the descriptions of symbols). Or equivalently, we suppose there are $n$ Reduce operations, and the number of pairs (load) for the $j$th operation is $k_j$. The schedule assigns exactly one Reduce task slot to each Reduce operation, and we use a set of binary variables to denote this: $x_{ij}$ ($i = 1, 2, …, m; j = 1, 2, … n$). $x_{ij} = 1$ indicates the $j$th Reduce operation is assigned to the $i$th Reduce task slot, and $x_{ij} = 0$ otherwise. Therefore, the total number of pairs to process on the $i$th task slot is:

$$p_i = \sum_{j=1}^{n} k_j x_{ij} \quad i = 1, 2...m$$

we call it the *load* of the $i$th Reduce task slot. One of the most widely used criteria for load balance is *max-load*, which is defined as:

$max\text{-}load(p_1, p_2, … , p_m) = max(p_1, p_2, … , p_m)$

Small max-load indicates balanced loads. With this criterion, the scheduling problem can be formulated as the following integer program:

min $p$

s.t. $p_i = \sum_{j=1}^{n} k_j x_{ij} \quad i = 1, 2...m$

$p_i \leq p \quad i = 1, 2,...,m$

$\sum_{i=1}^{m} x_{ij} = 1 \quad j = 1, 2,...,n$

$x_{ij} \in \{0,1\} \; i = 1, 2,...,m; \; j = 1, 2,...,n$

According to the standard notation for scheduling problems [G+79], this problem is denoted as P||$C_{max}$. It has been proved to be strongly NP-hard [Ho98]. This may be another reason for MapReduce to adopt the simple hash based strategy. To formulate this problem, we need the values of $k_1, k_2, … , k_n$. These values are collected by OS4M's communication mechanism, whose details will be described in the next section.

## 4 ENHANCEMENTS TO MAPREDUCE

This section discusses OS4M's extension to MapReduce. To formulate the scheduling problem, we need the key distribution of intermediate pairs, which is collected by the communication mechanism in Section 4.1. Our scheduling algorithm is briefly discussed in Section 4.2. For some jobs, the number of Reduce operations is extremely large incurring much network cost. This is solved by operation clustering described in Section 4.3. Finally, in Section 4.4, we discuss the Reduce pipelining technique, which increases resource utilization and decreases task durations.

*4.1 The Communication Mechanism*

According to the current MapReduce specification, different Map/Reduce operations are totally independent, without any communication between them. This works well for most scenarios, but for others, it is necessary to provide some communication mechanism to gather local statistics of each operation to evaluate some global statistics, and then let each operation take action according to the global statistics. According to the previous sections, the problem of scheduling Reduce operations involves collecting key distribution of pairs. So the above mechanism is required for scheduling Reduce operations. Although the global counter of Hadoop support collecting statistics from all operations and evaluating global information [Wh12], the collected statistics are not aggregated until the end of a MapReduce job, which is useless with respect to making decisions based on it. Besides, the global counter only supports integer data type, but a general application may require collecting other types of data.

We design a communication mechanism based on the Master-Slave architecture of MapReduce. According to the MapReduce specification, a daemon is installed on each cluster node. The daemon of the master is named JobTracker, which is responsible for maintaining states of slaves, scheduling tasks, etc. The daemon of each slave is named a TaskTracker, whose responsibility is to coordinate task slots on the cluster node, and assign tasks to them. Therefore, each task corresponds to exactly one TaskTracker. According to our communication mechanism, a task may communicate with its TaskTracker, and a TaskTracker may communicate with the JobTracker.

Given a MapReduce job, suppose there is a predetermined order of keys: $key_1, key_2, … key_n$ (Details of the ordering mechanism is discussed in Section 4.3). The communication mechanism works as follows:

1) Each Map operation sends its local statistics to its TaskTracker. In particular, the $i$th Map operation sends an $n$-dimensional vector to the TaskTracker:

$$K^{(i)} = (k_1^{(i)}, k_2^{(i)},...,k_n^{(i)})^T \qquad (4\text{-}1)$$

where $k_j^{(i)}$ is the number of pairs produced by the $i$th Map operation with key $key_j$.

2) The TaskTracker receives messages from Map operations on the local host, buffers them, and sends them to the JobTracker.

3) The JobTracker receives messages from TaskTrackers. When the statistics of a job are complete, it aggregates them to obtain the key distribution of intermediate pairs:

$$K = \sum_{i=1}^{M} K^{(i)} = (k_1, k_2,...,k_n)^T$$

$M$ is the total number of Map operations of the job.

4) So far, input of the scheduling problem is complete, so the JobTracker invokes our scheduling algorithm, and sends the resulted schedule to each TaskTracker. The message to each TaskTracker is an $n$-dimensional vector:

$$S = (s_1, s_2,...,s_n)^T$$

which means Reduce operation for key $key_j$ is assigned to the Reduce task with ID $s_j$.

5) Each TaskTracker forwards the schedule from the JobTracker to its local Reduce tasks.

6) After receiving the schedule from the TaskTracker, each Reduce task fetches data accordingly.

According to the above process, it can be seen that Reduce tasks cannot start copying Map outputs until all Map operations are finished and the schedule is received. Therefore, the I/O contention between Map and Reduce tasks no longer exists. This significantly accelerates Map tasks (See our experimental results in Section 5.1.2).

*4.2 The Operation Scheduling Algorithm*

As discussed in Section 3.2, the problem of scheduling Reduce operations is equivalent to P||$C_{max}$, which is strongly NP-hard. Existing algorithms for P||$C_{max}$ include heuristics and approximation algorithms. Graham introduced a 2-approximation algorithm [Gr66] and a 4/3 approximation algorithm [Gr69]. These two algorithms are simple and fast, but schedules produced by them are far from the optimal schedule. The approximation schemes introduced by Graham [Gr69] and Sahni [Sa76] may achieve any desired precision, but their time complexities are exponential in the number of task slots ($m$), so they are not applicable to large-scaled clusters. Hochbaum and Shmoys introduced a polynomial time approximation scheme for P||$C_{max}$ [HS87]. This scheme requires a long time to obtain a solution with high precision. For a solution within 1+ε of the optimal max-load, the scheme requires $O((n/\varepsilon)^{1/\varepsilon})$ time.

Therefore, existing algorithms can be divided into two types: one type includes heuristics that are fast but imprecise; the other type of algorithms can achieve desired precision, but the time complexity is large. Therefore, we design an algorithm that achieves the trade-off between high precision and low time complexity. Due to space constraint, we only give a brief discussion of the algorithm. Full details can be found in Section 5 of our manuscript [F+14]. We employ a technique named *dynamic programming decomposition* [F+14] to reduce the original problem into a series of weakly NP-hard sub-problem. Each sub-problem solves the problem of selecting Reduce operations for one Reduce task slot. We define the sub-problem the Balanced Subset Sum (BSS) problem. We design exact and approximation algorithms for the BSS problem. Our scheduling algorithm is based on the approximation algorithm for BSS.

*4.3 Operation Clustering*

The extensions of OS4M described above effectively supports collecting and aggregating statistics from all operations. However, it may also introduce performance cost. The cost can be divided into two classes. One is the computational overhead for calculating the schedule, and the other is the network overhead for collecting statistics and broadcasting schedule. According to our evaluations (see Fig. 10, Section 5.2.1), however, influence of the former is neglectable, so we focus on the network overhead.

In the collecting step, the network flow from each Map operation to its TaskTracker is 8$n$ bytes (we use the *long* type for $k_j^{(i)}$ in formula (4-1), whose width is 8 bytes in Java), so the total network flow from all Map operations to TaskTrackers is 8$Mn$ bytes. Similarly, the network flow from all TaskTrackers to the JobTracker is at most 8$Mn$ bytes (TaskTracker may combine statistics before sending them to the JobTracker). In the broadcasting step, the network flow from the JobTracker to each TaskTracker is 4$n$ bytes (we use the *int* type to represent the schedule element, whose width is 4 bytes), so is the network flow from each TaskTracker to each Reduce task. Therefore, the total network flow from the JobTracker to all TaskTrackers is 4$tn$, where $t$ is the number of TaskTrackers (see Table 1). The total network flow from all TaskTrackers to all Reduce tasks is 4$rn$, where $r$ is the number of Reduce tasks. In summary, the network flow in the collecting step is at most 16$Mn$, and the network flow in the broadcasting step is 4$n(t + r)$. The total network flow is at most 4$n(4M + t + r)$.

The analysis above indicates that if the number of Reduce operations ($n$) is large, there will be large network overhead. To reduce the overhead, we reduce the number of Reduce operations by combining a set of Reduce operations into an *operation cluster*. The operation cluster will be treated as a single unit for scheduling. In addition, each operation cluster is assigned a unique integer ID, which can be used in ordering the Reduce operation clusters.

OS4M leaves API for users to employ their customized clustering algorithm. If no clustering algorithm is specified, OS4M adopts the default clustering algorithm, which requires only one parameter: the targeted number of Reduce operations $n$. According to the default algorithm, operations with keys $key_i$ and $key_j$ are in the same cluster, if and only if

$$|Hash(key_i)| \equiv |Hash(key_j)| \,(\mathrm{mod}\, n)$$

Their cluster ID is $|Hash(key_i)|$ (mod $n$) + 1. It can be verified that this algorithm is self-adaptive in the sense that the number of Reduce operation clusters is at most the targeted number of Reduce operations.

*4.4 Reduce Pipelining*

Compared with default MapReduce, Reduce tasks of OS4M cannot copy Map outputs until all Map operations are finished. This makes Map operations faster. However, it also causes some delay in Reduce tasks. We solve this problem by introducing a mechanism named Reduce pipelining.

Before giving details, we briefly review the default workflow of a Reduce task. Each Reduce task goes through three phases (See Fig. 4(a)): 1) *Copy*. In this phase, a Reduce task initiates a number of threads to fetch data from Map operations. The data are transferred through HTTP GET; therefore, the bottleneck resource of this phase is network bandwidth. 2) *Sort*. When input data for all operations of this task are fetched, they are sorted by key so that pairs of the same operation are grouped together. The sort process is usually divided into multiple passes, with the output of one pass being the input of the next pass. For jobs with big data, the input and output data of a pass are usually stored in local disk. Therefore, this phase is disk I/O intensive. 3) *Run*. Each operation is processed by one invocation of the Reduce function. Depending on the specific implementation, the Reduce function may have different bottleneck resources. According to our experience, most Reduce functions in

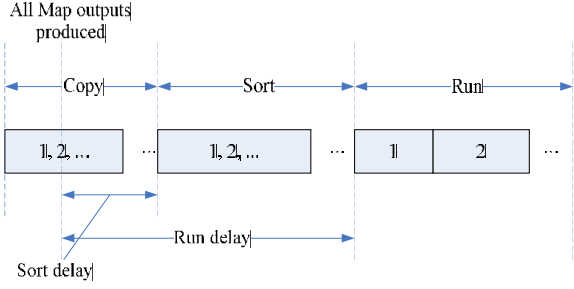

(a) Reduce task structure for default MapReduce

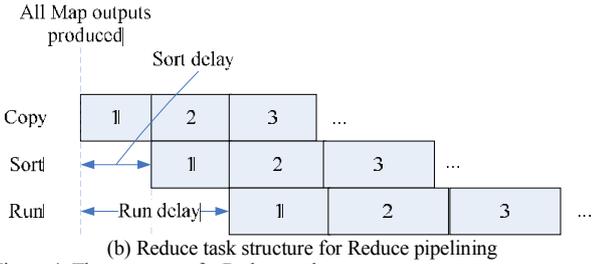

(b) Reduce task structure for Reduce pipelining

Figure 4. The structure of a Reduce task

practice (like the benchmarks in PUMA [T+12]) are CPU intensive. It can be seen that different phases mainly consumes different resources. However, they are processed in sequence by default MapReduce. That is, a phase cannot start until the previous phase is finished for all operations of this task. This will cause significant waste of resources.

To solve the above problems, we design a pipeline to parallelize the three phases (see Fig. 4(b)). The general idea of Reduce pipelining is as follows: Suppose a Reduce task has more than one operation[2]. First, the task copies input of the first operation and then sorts and runs it. When sorting the first operation, it simultaneously copies input of the second operation. After that, the task sorts the second operation, and simultaneously copies input for the third operation and runs the first operation, and so on.

Through this mechanism, the input of a Reduce task is split into many small parts, which are processed separately. The split is based on the Reduce operation. That means input of a Reduce operation must be in the same part. Therefore, the split does not violate the Reduce Input Constraint, and the correctness of the Reduce task is guaranteed. Splitting the input into small parts has another benefit: the small parts have a better chance of being sorted in memory than on disk. The former is much faster than the latter. To facilitate fetching Map outputs, the Map output corresponding to each operation (cluster) is written as a separate bucket file. So when a Reduce operation's input is requested, the TaskTracker does not need to seek the part from a large file, or transfer the whole file. This further improves the performance.

To guarantee the correctness of a MapReduce job, there is an inherent barrier between Map and Reduce phases

---

[2] For this and the following discussions, the term operation may also indicate an operation cluster, depending on whether the user adopts an operation clustering algorithm.

TABLE 2. BENCHMARKS USED IN THE EVALUATIONS

| Name | Abbreviation | Brief description |
|---|---|---|
| Adjacent-list | AL | Generate the adjacent/reverse adjacent lists of a graph used by a Page-Rank like algorithm |
| Inverted-index | II | Build a word-to-document index given a set of documents. |
| Ranked inverted index | RII | Given a list of words with their frequencies and documents, produce the list of documents containing the word in decreasing order. |
| Sequence Count | SC | Given a set of documents, counts the occurrences of unique three-consecutive words per document. |
| Self-join | SJ | Given a set of $k$-field associations, generate $k+1$ field associations. |
| Term Vector | TV | Determine the frequent words in each host. |

[L+11]: the Reduce function cannot be called until the last Map operation is finished. To minimize this barrier, Hadoop and OS4M adopt different approaches: Hadoop achieves this by overlapping Map tasks and the copy phase of Reduce tasks, whereas OS4M does this by Reduce pipelining. To measure the barrier, we define the *sort delay* as the duration from when all Map outputs are produced (when all Map operations are finished) to when the first Reduce operation enters the sort phase (see Fig. 4). Similarly, we define the *run delay* as the duration from when all Map outputs are produced to when the first Reduce operation enters the run phase. Experimental results for these delays will be given in Section 5.2.3. *To minimize these delays, we sort operations in the pipeline by the increasing order of their loads.*

## 5 EVALUATIONS

We have implemented OS4M and transparently incorporated it to MapReduce. The users do not need to change their MapReduce code; they only need to replace our jar library with the default library. Optionally, the users may set the targeted number of Reduce operations. We chose Hadoop 1.0.4 as the code base for OS4M and as the baseline for comparison. The reasons are two fold: 1) It is a stable version. 2) Hadoop 1.x is the most widely used in production clusters [Wh12]. We evaluate OS4M and compare it to MapReduce (Hadoop 1.0.4) by Purdue MapReduce Benchmarks Suite (PUMA) [T+12], one of the most widely used MapReduce benchmark suites. The benchmarks used in our evaluations and their brief descriptions are listed in Table 2. More detailed descriptions can be found in [T+12]. Each benchmark runs on three datasets: the largest is denoted by L, the medium is denoted by M, and the smallest is denoted by S. So II_S refers to Inverted Index on the smallest dataset, and TV_L refers to Term-Vector on the largest dataset, and so on. The sizes and sources of the benchmarks' input data are given in Table 3.

All experiments are conducted on a homogeneous cluster with 9 VMs on the IBM RC2 Cloud platform [A+10]. One VM runs the JobTracker and NameNode, while each of the other 8 runs a TaskTracker and a DataNode. Each VM has 4 virtual CPU of 2.93GHz, 4GB memory and Fedora 14 operating system. According to our test, the bandwidths for network, disk read, and disk write are 37 MB/s, 203 MB/s, and 121 MB/s, respectively. Unless otherwise stated, we use default Hadoop parameters, with the following exceptions:

1) The number of Reduce tasks is set to 0.95* <number of VMs> * mapreduce.tasktracker.reduce.tasks.maximum = 0.95 * 8 * 4 ≈ 30, as recommended by Apache Software Foundation [Ma13].

2) For OS4M, if there are more than 240 Reduce operations, we use the default clustering algorithm described in Section 4.3 to combine Reduce operations.

3) We place 4 Map task slots and 4 Reduce task slots on each node, as recommended by the original MapReduce proposal [DG04].

4) To make good use of memory, we set the JVM heap space of each Map/Reduce task to be 500MB.

5) For OS4M, the η parameter of the scheduling algorithm is set to 0.002, so the relative error is at most 0.2% (see [F+14]).

TABLE 3. FEATURES OF BENCHMARKS' INPUT DATA

| Benchmark | Data source | Data Size (GB) | | |
|---|---|---|---|---|
| | | Small | Medium | Large |
| AL | From PUMA website | 5 | 10 | 15 |
| II | Wikipedia dump files | 5 | 10 | 15 |
| RII | Output of SC | 10 | 20 | 30 |
| SC | Wikipedia dump files | 5 | 10 | 15 |
| SJ | From PUMA website | 10 | 20 | 30 |
| TV | Wikipedia dump files | 5 | 10 | 15 |

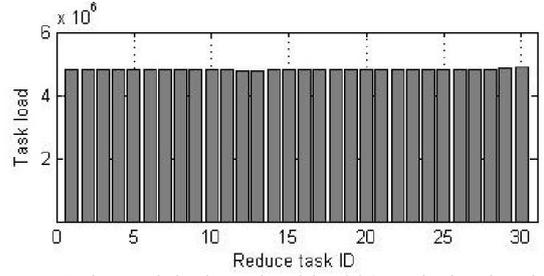

Figure 5. Reduce task loads produced by OS4M. The benchmark, input data, and cluster configuration are identical to that of Fig. 1.

Finally, note that each result is obtained by running 3 jobs and choosing the one with the smallest duration.

### 5.1 Benefits Introduced by OS4M

OS4M brings two major benefits: better load balance, and shorter task duration.

#### 5.1.1 Better Load Balance

In Section 1, we give the Reduce task loads of RII_S produced by default MapReduce. Fig. 5 shows the results produced by OS4M with the same benchmark, input data, and cluster configuration. By comparing Fig. 5 and Fig. 1(b), it can be found that OS4M achieves better load balance.

Load balance is not only achieved for RII_S. To demonstrate this, Fig. 6 shows the max-load divided by the ideal load, which is the load when all Reduce tasks share equal load:

$$p_{ideal} = \frac{1}{r}\sum_{i=1}^{n} k_j$$

It can be proved that $p_{ideal}$ is a theoretical lower bound of the optimal max-load, so the values in Fig. 6 are at least 1. Recall from Section 3.2 that smaller max-load indicates better load balance. Therefore, the closer a value in Fig. 6 is to 1, the better load balance is achieved.

According to the above discussion, it can be observed that for all cases, the max-loads produced by OS4M are smaller, compared with those produced by Hadoop. This means better load balance is achieved by OS4M. In addition, the standard deviations of Reduce task loads (relative to mean load) produced by OS4M are much smaller. This is another indicator of better load balance. For all benchmarks but II, the max-loads produced by OS4M divided by ideal loads are close to 1. Therefore the schedules produced by our scheduling algorithm are close to the optimal schedules.

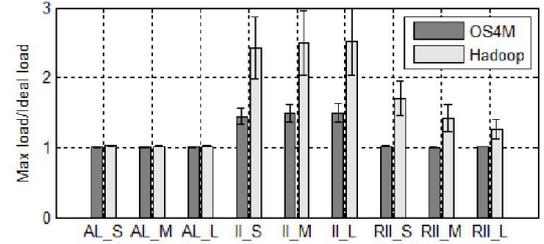

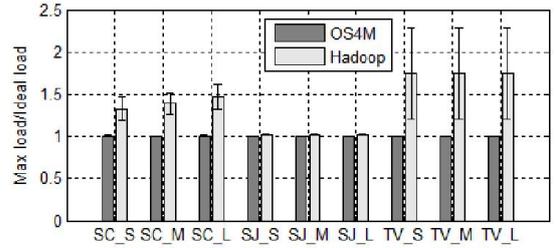

Figure 6. Max-load of Reduce tasks divided by the ideal load. The error bars indicates standard deviactiation divided by mean.

#### 5.1.2 Shorter Task Duration

Fig. 7 shows the average Reduce task duration, and their standard deviation divided by the average. It can be seen that for all cases, the average Reduce task durations produced by OS4M are smaller than those of Hadoop. The reasons are twofold: 1) Due to Reduce pipelining, the three phases of a Reduce task is parallelized to some extent, which makes OS4M faster. 2) For Hadoop, Reduce tasks start to copy input after the first Map operation is over. For OS4M, the Reduce tasks cannot begin until all Map operations are finished. So Reduce tasks of Hadoop start much earlier. In Fig. 7, for 16 of the 18 cases, the standard deviations produced by OS4M are smaller, indicating more balanced task durations for OS4M. This is due to the load balance achieved by our scheduling algorithm.

We next evaluate Map task durations, as shown in Fig. 8. It can be seen that, for all cases, the average Map task durations produced by OS4M are much smaller than those of Hadoop. According to Section 4.1, this is because OS4M removes the I/O contention between Map and Reduce tasks.

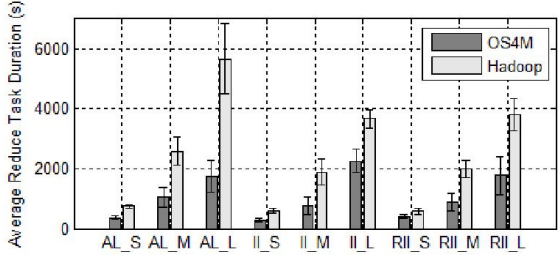
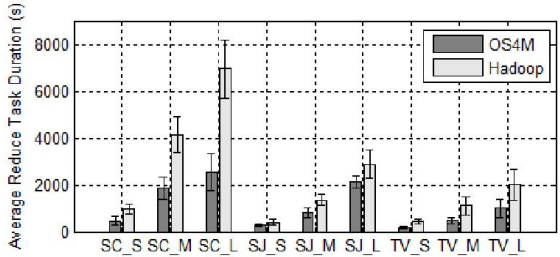
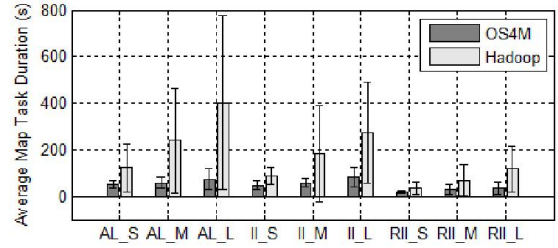
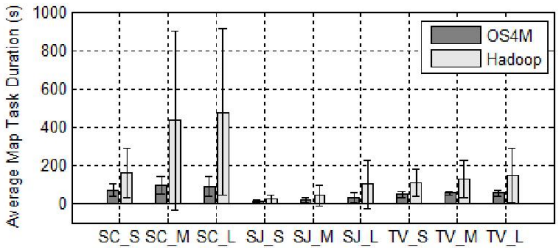

Figure 7. Average Reduce task duration. The error bars indicates standard deviation divided by mean.

Figure 8. Average Map task duration. The error bars indicates standard deviaction divided by mean.

We verify this by the progress plot of II_S, which we have used in Section 1 to demonstrate the problem. Recall that the input size for II_S is 5GB (see Table 3), which gives rise to 80 Map tasks (because the default HDFS block size is 64 MB). Since there are 32 Map slots, the Map tasks are divided into 3 waves, with 32, 32 and 16 tasks, and corresponding to 40%, 80% and 100% progress of the Map phase, respectively (see Fig. 9). Before the end of the first wave, no Map task is finished, so there is no I/O contention for Hadoop. In this wave, the progress rates of OS4M and Hadoop are almost identical. In the second wave, as some Map outputs are generated, the I/O contention of Hadoop starts to emerge, so Hadoop progresses slower than OS4M. In the last wave, as more Map outputs are generated, the I/O contention of Hadoop becomes even more intensive. This further slows down its progress. For OS4M, the progress rate of the 3 waves remains almost consistent. This can be further verified by Fig. 8: the standard deviations produced by OS4M are much smaller, compared with Hadoop.

### 5.2 Costs Introduced by OS4M

OS4M introduces three costs: 1) Compared with default MapReduce, OS4M involves a centralized scheduling algorithm, which will take some extra time. 2) The network overhead of collecting operation statistics and broadcasting the schedule. 3) For OS4M, Reduce tasks cannot begin until all Map operations are finished. This may cause some delay compared with default MapReduce. We evaluate these costs separately.

#### 5.2.1 Time Spent on the Scheduling Algorithm

The time spent on our key distribution based operation scheduling algorithm is shown in Fig. 10. For all cases, the running times are smaller than 0.5 second, which is trivial compared with the job duration. In addition, it can be observed that for each benchmark, the time spent on scheduling the largest dataset is close to that spent on the smallest dataset. For example, the time spent on AL_S is 0.097 second, and the time on AL_L is 0.091 second. This indicates our operation scheduling algorithm is scalable.

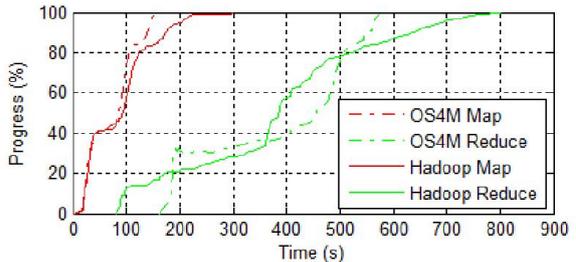

Figure 9. Progress plot of Map/Reduce phases for II_S. The configuration is identical to that in Fig. 2.

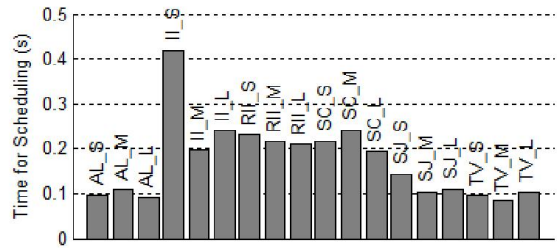

Figure 10. Running time of the scheduling algorithm (in second).

#### 5.2.2 Network Overhead

OS4M's network overhead comes from two aspects: 1) Collecting Map operations' statistics. 2) Broadcasting the schedule to Reduce tasks. Fig. 11 gives these overheads for all cases. It can be seen that the overhead for collecting statistics is much greater than the overhead for broadcasting schedule. For all cases, the total network flow is smaller than 2MB. This amount of data is trivial compared with those of the input/shuffle/output data. Therefore, it can be concluded that the network overhead incurred by OS4M is insignificant.

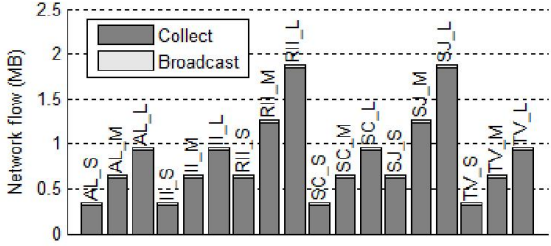

Figure 11. Network overhead of OS4M (in MB).

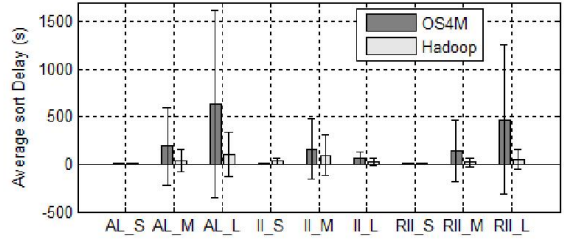
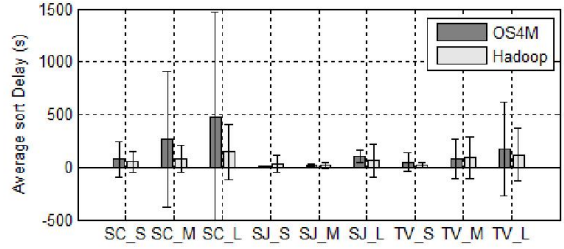

Figure 12. Average sort delay (in second). The error bars indicate standard deviation divided by the average.

*5.2.3 Delays Caused by OS4M*

As discussed in Section 4.4, there is an inherent barrier between Map and Reduce tasks. We measure this barrier by the sort delay and run delay (see Fig. 4). We show the average sort delay in Fig. 12. For most cases, OS4M's sort delays are larger than Hadoop's. This can be expected because Hadoop starts fetching data much earlier than OS4M. However, for 4 cases (AL_S, II_S, SJ_S and TV_M), OS4M's sort delays are shorter. This is explained as follows: through Reduce pipelining, the input of a Reduce task is divided into many small parts, each for an operation (cluster). So the size of each part is much smaller than the whole input, and copying the first part by OS4M can be finished earlier than copying the whole input by Hadoop, even though the latter starts much earlier.

Similar analysis can be made for run delays, which are shown in Fig. 13. In particular, the average run delays produced by OS4M are smaller for 5 of the 18 cases (AL_S, II_S, SJ_S, TV_M, and RII_S). For 7 of the remaining 13 cases, the differences of average run delays is less than one minute, which is insignificant compared their job durations (see Section 5.3). This is caused by two factors: 1) Through Reduce pipelining, OS4M each time copies a small part of Reduce input, so it sorts faster than Hadoop, which sorts the whole input. 2) Since OS4M processes small fractions, it is more likely that the fraction will be stored and sorted in memory, which is much faster than storing and sorting it on disk.

It can be observed that for small input data, OS4M's sort and run delays are more likely to be smaller Hadoop's. This is because small input data leads to fewer waves of Map operations, so that OS4M's Reduce tasks begin to fetch data not too much later compared with Hadoop.

*5.3 The Overall Effects*

All the benefits and costs discussed above are reflected and integrated in the job duration. In Fig. 14, we give the job duration of OS4M divided by the job duration of Hadoop. In addition, the job durations for Hadoop are given in Table 4 for reference. For all cases, the job durations produced by OS4M are smaller than Hadoop. This is expectable since from the results above, it can be seen that the cost introduced by OS4M is trivial compared with the benefits. From Fig. 14 it can be seen that for 14 of the 18 cases, the performance increase is at least 20%. The largest performance increase is achieved by AL_L (42%), while the smallest increase is achieved by SJ_L (8%).

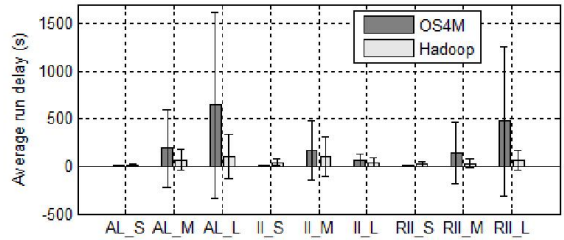
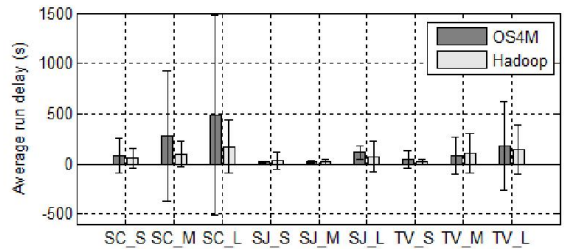

Figure 13. Average run delay (in second). The error bars indicate standard deviation divided by the average.

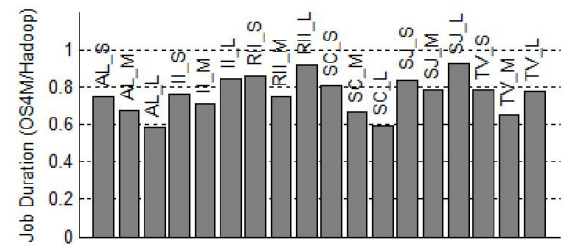

Figure 14. Job duration of OS4M divided by the job duration of Hadoop

TABLE 4. JOB DURATIONS FOR HADOOP

| Benchmark | Job duration (s) | | |
| --- | --- | --- | --- |
| | Small | Medium | Large |
| AL | 943 | 3002 | 6417 |
| II | 813 | 2388 | 4213 |
| RII | 819 | 2363 | 4267 |
| SC | 1299 | 4721 | 7787 |
| SJ | 597 | 1717 | 3446 |
| TV | 662 | 1726 | 2764 |

In summary, OS4M introduces both performance costs and benefits. However, the costs are insignificant compared with the benefits. Therefore, OS4M significantly improves the performance of MapReduce jobs.

### 5.4 Parameter Sensitivity Evaluation

OS4M allows the user to specify the targeted number of Reduce operations (see Section 4.3). This parameter may have performance implications: If the value is small, the granularity of Reduce operation cluster will be large, which makes it difficult to achieve load balance. On the other hand, a large parameter value will result in too many Reduce operation clusters, incurring much cost for starting threads, writing outputs, etc. In this section, we evaluate the performance implications of this parameter.

To avoid empty Reduce operations, we use synthesized benchmark and dataset. The dataset contains 7GB positive random integers uniformly distributed between 1 and $10^6$, and the benchmark calculates their histogram. We set the hash value of each integer to itself. Since the integers are uniformly distributed, there is no problem of load balance. This allows us to focus on other performance factors.

Fig. 15 shows the average time spent on each of the three phases in Reduce pipeline. When the number of operation clusters is small (< 180), the sort time is long, because each operation cluster is large, and cannot be sorted in memory. When the number of clusters is large (> 480), the three average times all increase. This is because the granularity of Reduce pipeline is too fine, causing other performance costs (starting threads, writing outputs, etc.) to increase. It can be seen that the three average times are shortest when the targeted number of Reduce operations is between 180 and 480. That is, when the number of Reduce operation clusters is between 6 and 16 times the number of Reduce tasks. In this range, the performance is relatively insensitive to the parameter value. It should be noted that this a wide range. In addition, the performance costs increase mildly outside this range. Therefore, OS4M does not require much tuning effort to achieve good performance.

### 5.5 Scalability Evaluation

To evaluate the scalability of OS4M, we run OS4M and Hadoop on different numbers of cluster nodes. We use the TV benchmark, and a 12 GB Wikipedia dump file as input.

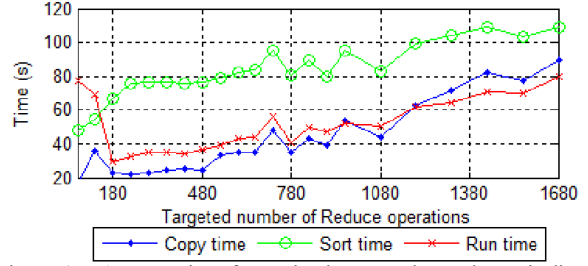
Figure 15. Average time for each phase on the Reduce pipeline (in second). The benchmark and data are synthesized.

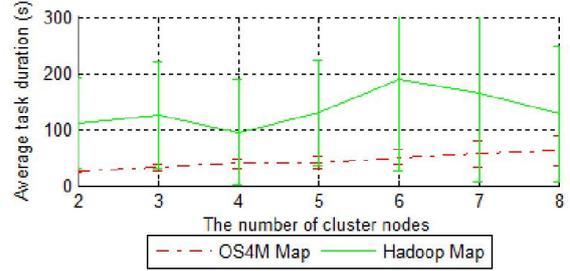
(a) Average Map task duration

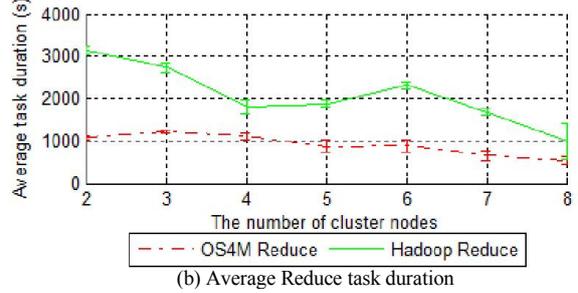
(b) Average Reduce task duration

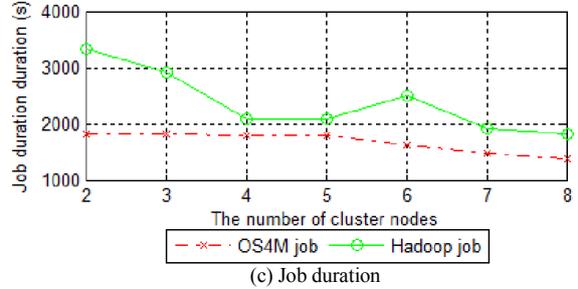
(c) Job duration

Figure 16. Results on different numbers of cluster nodes. The benchmark is TV and input is a 12GB Wikipedia dump file. Error bars indicate standard deviation divided by mean.

For each job, we use all the Reduce task slots, so the number of Reduce tasks is 4 times the number of nodes. The average Map task durations, average Reduce task durations, and job durations are given in Fig. 16. For all cases, the 3 durations produced by OS4M are all smaller. The performance gain of OS4M is most notable when the number of nodes is small. In particular, the performance gain is 46% on 2 nodes. This is because with fewer nodes, the data for each node is larger, leading to more waves of Map operations. Therefore, Hadoop's I/O contention between Map and Reduce tasks is more intensive. So it can be expected that the performance gain of OS4M will be greater on larger input data.

## 6 DISCUSSIONS

Fault-tolerance is a fundamental feature of MapReduce. It promotes successful jobs in the presence of node and task failures. We argue that OS4M's communication mechanism (described in Section 4.1) supports MapReduce's build-in fault tolerance mechanism. In particular, the JobTracker maintains a hash map to store statistics of Map tasks. The key of the hash map is the Map task ID, and the value is its statistics. Whenever a Map task attempt is finished, it sends its statistics to the TaskTracker. The TaskTracker checks the status of the task attempt. If the task attempt is finished successfully, the TaskTracker sends the statistics to the JobTracker through the next heartbeat message; otherwise, the statistics are discarded. When the JobTracker receives the statistics, it puts the task ID and statistics in the hash map. Therefore, it can be seen that: 1) If a node and its TaskTracker fails, the JobTracker will detect this and assign its tasks to other TaskTrackers. The task ID of the newly assigned tasks will be unchanged. Therefore, when the newly assigned tasks finishes, the JobTracker will correctly update the hash map. 2) If a Map task is re-executed or speculatively executed, it has multiple task attempts. Although these attempts have different attempt IDs, they share the same task ID. Therefore, no matter how many attempts a task may have, there is exactly one entry in the JobTracker's hash table for it. This entry corresponds to a successful attempt of the task.

A Reduce operation cluster may contain more than one Reduce key, so it is similar to a Reduce task in this sense. However, they are significantly different in the following aspects: 1) An operation cluster is a much smaller unit compared with a task. Recall from Section 5.4 that we recommend the number of Reduce operation clusters to be 6 to 16 times the number of Reduce tasks. So the granularity of a Reduce operation cluster is much smaller, which makes it easier to achieve load balance. 2) A Reduce operation cluster is much less resource-intensive than a Reduce task. To achieve load balance, one can either increase the number of Reduce tasks, or increase the number of Reduce operation clusters. The former will introduce significant performance penalty [A+12], whereas the latter only introduces mild performance cost (see Fig. 15). 3) The users may customize the operation clustering algorithm to fulfill their specific requirements.

One may argue that the data used in our evaluations are not large enough (from 5GB to 30GB, see Table 3). We leverage two results to show that the scales of our evaluations are typical in practice: 1) Ren et al. have analyzed the job traces of clusters in CMU Parallel Data Lab. The job traces involve more than 19,000 jobs from different departments of CMU for more than 20 months. The analysis shows that more than 50% of jobs touch less than 10 GB data (This represents the total size of input, shuffle, and output data of a job). In addition, for more than 50% of jobs, the job durations are less than 640 seconds [R+13]. 2) According to job trace analysis from Google clusters [DG08], the average job durations are 634 seconds in Aug. 2004, 874 seconds in Mar. 2006, and 395 seconds in Sept. 2007. Our job durations are no less than these durations.

## 7 RELATED WORK

**Scheduling MapReduce workload at the job level.** Tian et al. introduced a dynamic scheduler which classifies workloads into I/O and CPU-bound jobs, and schedules them through separate queues [T+09]. Kang et al. introduced a scheduling scheme for scenarios with multiple clusters and multiple jobs [K+11]. By batching I/O requests and reducing context switching, the scheme improves performance and enhances fairness between jobs. Sandholm and Lai introduced a dynamic priority scheduler for MapReduce [SL10], which allows the users to dynamically control their allocated capacity. Moseley et al. model MapReduce workload through the classic two-stage flow shop problem, and design a 12-approximation algorithm to minimize the total flow time of multiple jobs [M+11]. A novel framework named SkewReduce [K+10] transforms a feature-extraction application into a graph of MapReduce jobs, and eliminates skews for these jobs. Although these works improve the scheduling of MapReduce workload, they are job-level scheduling mechanisms. In other words, they regard a MapReduce job as the basic unit, and try to optimize some criterion in the presence of multiple jobs. Therefore, they are totally different from our work.

**Scheduling MapReduce workload at the task level.** Berlińska and Drozdowski model the MapReduce workload through the Divisible load theory [B+05], and design algorithms to partition and schedule them [BD11]. The focus of their work is on Map tasks, because they assume Reduce tasks have roughly equal execution times. In this study, we find it far from reality and instead focus on scheduling Reduce workload. Zaharia et al.'s algorithm improves performance by managing speculative tasks [Z+08]. It moves the task backup mechanism to heterogeneous clusters. Mao et al. introduced a task-level scheduler of MapReduce, which dynamically adjusts the task slots of cluster nodes [M+11b]. Tarazu [A+12] introduced by Ahmad et al. achieves task-level load balance on heterogeneous clusters by dividing jobs into shuffle-critical or Map-critical classes, computing processing rates and creating skewed tasks. The above works improve the scheduling of MapReduce at the task level, rather than at the operation level, so they are totally different from our work.

SkewTune introduced by Kwon et al. is an enhancement of MapReduce's task backup mechanism [K+12]. When a task slot is busy while another is free, part of the busy slot's workload is migrated to the free slot. Therefore, SkewTune improves the scheduling of MapReduce at the sub-task level. However, SkewTune is significantly different from OS4M in the following aspects: 1) *The overall method*. SkewTune adopts a local-search based method. Given a job, it first adopts the default schedule (a local optimum) generated by MapReduce. At runtime, when load imbalance is detected, the load of the straggler is repartitioned and migrated to other idle task slots (improving the local optimum). In contrast, OS4M generates only one (near) optimal schedule based on global information. 2) *The applicable scenarios*. SkewTune is a heavy-weight approach compared with OS4M. Each workload migration of SkewTune incurs a cost on the order of 30 seconds [K+12]. According to [DG08], the average job duration in Sept. 2007 is 395 seconds. Therefore, SkewTune is mainly applicable to long jobs. In addition, SkewTune incorporate extra components, like ST JobTracker and ST TaskTrackers, which make it more heavy-weight. In comparison, OS4M is a light-weight solution. It introduces little performance cost. Therefore, it is applicable to both long and short jobs.

**Pipelining MapReduce workload**. MapReduce Online improves the connection between Map and Reduce tasks [C+10]. It sends Map task outputs to Reduce tasks immediately after they are produced, without being materialized to local files first. This essentially puts Map and Reduce tasks on a pipeline, thereby reducing the job duration. However, MapReduce Online is significantly difference from OS4M, because for MapReduce Online, the pipeline is between Map and Reduce tasks, whereas for OS4M, the pipeline is within Reduce tasks.

Hadoop-A introduced by Wang et al. leaves Map outputs on remote disk until it is time to merge them [W+11]. It also introduces a pipeline in the Reduce task. Hadoop-A is significantly different from OS4M because: 1) Hadoop-A's methodology depends on the assumption that network and memory have similar performance. So it requires RDMA (Remote Direct Memory Access) hardware support. 2) Hadoop-A does not support configuring the granularity of the pipeline. This may have significant performance implications (see Section 5.4). 3) Hadoop-A does not support configuring the splits' ordering on the pipeline, which may affect the barrier between Map and Reduce tasks (see Section 4.4).

Li et al. design a novel hash-based platform to enable fast in-memory process [L+11]. It introduces overlaps among the three phases of Reduce tasks. However, since it assumes that the Reduce function is commutative and associative, it is not applicable to general MapReduce jobs.

**Other Improvements to MapReduce**. Gufler et al. introduced a mechanism to collect and aggregate statistics for MapReduce workload [G+12]. When the statistics size is large, it employs algorithms to approximate the global statistics by neglecting small operations, and assuming they have uniform distribution. So it is different from OS4M: Even if an operation clustering is adopted, the total load for each operation cluster is exact, rather than approximated. This is important for our scheduling algorithm to produce globally (near) optimal schedules.

## 8 CONCLUSIONS

MapReduce is one of the most important frameworks in distributed and Cloud computing. A critical problem for it is workload scheduling. Traditional methods schedule workload at coarse-grained levels, such as job level, or task level. In this study, we schedule MapReduce workload at a fine-grained level: operation level.

In this paper, we introduce a set of mechanisms named OS4M to achieve load balance by scheduling at the operation level. OS4M works by collecting Map operation statistics through our communication mechanism, and scheduling Reduce operations based on the key distribution of intermediate pairs. For OS4M, the overlap between Map and Reduce tasks is removed, which significantly accelerate Map tasks. In addition, OS4M places the 3 phases of Reduce tasks into a pipeline to minimize the barrier between Map and Reduce tasks.

Experiments on PUMA benchmarks show that OS4M achieves load balance, and shortens task durations. Although it also introduces performance costs, they are trivial compared with the benefits. Compared with Hadoop, OS4M increases performance by up to 42%. In the future, we will move OS4M to the heterogeneous cluster and heterogeneous task slots.